\documentclass[aps,prl,twocolumn,superscriptaddress,10pt]{revtex4-1}
\usepackage{graphicx}
\usepackage{color}
\usepackage{multirow}
\usepackage{amsmath}
\usepackage{amssymb}
\usepackage{bm}
\usepackage{placeins}

\makeatletter
\newcommand*{\balancecolsandclearpage}{%
  \close@column@grid
  \clearpage
  \twocolumngrid
}
\makeatother

\renewcommand\vec[1]{\overrightarrow{#1}}
\newcommand\cev[1]{\overleftarrow{#1}}

\begin{document}
\title{First-principles calculation of spin-orbit torque in a Co/Pt bilayer}

\author{K. D. Belashchenko}
\affiliation{Department of Physics and Astronomy and Nebraska Center for Materials and Nanoscience, University of Nebraska-Lincoln, Lincoln, Nebraska 68588, USA}

\author{Alexey A. Kovalev}
\affiliation{Department of Physics and Astronomy and Nebraska Center for Materials and Nanoscience, University of Nebraska-Lincoln, Lincoln, Nebraska 68588, USA}

\author{M. van Schilfgaarde}
\affiliation{Department of Physics, Kings College London, Strand, London WC2R 2LS, United Kingdom}

\date{\today}

\begin{abstract}
The angular dependence of spin-orbit torque in a disordered Co/Pt bilayer is calculated using a first-principles non-equilibrium Green's function formalism with an explicit supercell averaging over Anderson disorder. In addition to the usual dampinglike and fieldlike terms, the odd torque contains a sizeable planar Hall-like term $(\mathbf{m\cdot E})\mathbf{m}\times(\mathbf{z}\times\mathbf{m})$ whose contribution to current-induced damping is consistent with experimental observations. The dampinglike and planar Hall-like torquances depend weakly on disorder strength, while the fieldlike torquance declines with increasing disorder.
The torques that contribute to damping are almost entirely due to spin-orbit coupling on the Pt atoms, but the fieldlike torque does not require it.
\end{abstract}

\maketitle

Spin-orbit torque (SOT) \cite{Manchon.Miron.eaAe2018}, which is a manifestation of relativistic physics in solid-state systems, has attracted considerable interest due to its device applications \cite{Ramaswamy.Lee.eaAPR2018} in memory technologies \cite{Garello.Avci.eaAPL2014,Yang.Ryu.eaNN2015,Aradhya.Rowlands.eaNL2016,Prenat.Jabeur.eaITMCS2016,Fukami.OhnoJJAP2017} and spin-torque nano-oscillators \cite{Liu.Pai.eaPRL2012,Demidov.Urazhdin.eaNM2012,Duan.Smith.eaNC2014,Collet.Milly.eaNC2016,Awad.Duerrenfeld.eaNP2017}. SOT can arise in systems lacking bulk inversion symmetry, such as (Ga,Mn)As crystalline systems \cite{Kurebayashi.Sinova.eaNN2014}, or in systems lacking structural inversion symmetry. It can be described in terms of the nonequilibrium spin density  \cite{Garate.MacDonaldPRB2009,Manchon.ZhangPRB2009,Freimuth.Bluegel.eaPRB2014} and can affect the magnetization dynamics \cite{Ando.Takahashi.eaPRL2008}. For systems containing a heavy metal/ferromagnet interface, two mechanisms of SOT have been suggested: the inverse spin-galvanic effect (ISGE) \cite{LyandaGeller,EdelsteinSSC1990,Ganichev} arising at a heavy-metal/ferromagnet interface \cite{Chernyshov.Overby.eaNP2009,MihaiMiron.Gaudin.eaNM2010,Miron.Garello.ea2011,Manchon.Koo.eaNM2015,Fan.Celik.eaNC2014} and the bulk spin-Hall effect \cite{Sinova.Valenzuela.eaRMP2015} originating in the bulk of the heavy metal \cite{Liu.Moriyama.eaPRL2011,Liu.Pai.eaS2012,Liu.Lee.eaPRL2012}. These mechanisms lead to the fieldlike $(\mathbf{z}\times\mathbf{E})\times\mathbf{m}$ and dampinglike $\mathbf{m} \times [(\mathbf{z} \times \mathbf{E})\times \mathbf{m}]$ terms in SOT, which are, respectively, odd and even with respect to the magnetization described by the unit vector $\mathbf{m}$. Other terms with more complicated angular dependence are allowed by symmetry and have been experimentally identified in several systems \cite{Garello.Miron.eaNN2013,Ghosh2017,Safranski.Montoya.eaAe2017}. Such contributions can arise due to interfacial scattering \cite{Manchon2013}, even without any bulk spin-Hall effect \cite{Amin.StilesPRB2016,Amin.StilesPRB2016a}, and they are sensitive to the treatment of disorder \cite{Manchon2013,Qaiumzadeh.Duine.eaPRB2015,Ado.Tretiakov.eaPRB2017,XiaoNiu}. Axially asymmetric contributions to SOT induced by low crystalline symmetry have also been observed \cite{MacNeill.Stiehl.eaNP2017}.

The layers in SOT bilayers are usually made about a nanometer thick or even less. The phenomenological notion of an interface between bulk regions, as well as the interpretation in terms of the bulk spin-Hall effect, is, therefore, unjustified, and a fully quantum-mechanical treatment of the whole device is essential. An extreme case is that of a magnetic layer in contact with a topological insulator (TI) \cite{Mahfouzi.Nikolic.eaPRB2016,Hanke.Freimuth.eaNC2017}, which can generate strong SOT \cite{Mellnik.Lee.eaN2014}. There is ample experimental evidence of the existence of an interfacial contribution to SOT \cite{Ohno,Yang,Ohno1}.
\textit{Ab-initio} studies of Pt/Py bilayers also suggest the importance of interfacial contributions to the spin-Hall effect \cite{Kelly}, which should lead to an interfacial SOT.

Most of the existing \textit{ab-initio} studies of SOT rely on the use of phenomenological broadening for the Green's functions \cite{Freimuth.Bluegel.eaPRB2014,Mahfouzi2018}, which does not capture the full physics of SOT. A calculation of SOT using the coherent potential approximation (CPA) for disorder averaging was also reported \cite{Wimmer}, but only one orientation of the magnetization was considered.

In this Letter, we develop the non-equilibrium Green's function (NEGF) approach \cite{Faleev} within the tight-binding linear muffin-tin orbital (LMTO) method \cite{LMTO} for \emph{ab-initio} calculations of SOT in magnetic multilayered systems with explicit treatment of disorder and apply it to study SOT in a Co/Pt bilayer. Our results reveal a complicated angular dependence of SOT, including a sizeable planar Hall-like contribution.

In our LMTO-NEGF treatment, spin-orbit coupling is included as a perturbation to the second-order LMTO potential parameters \cite{Turek2008,Belashchenko2015}. The spin torque on atom $i$ is calculated as $\mathbf{T}_i=\int\mathbf{B}_{xc,in}(\mathbf{r})\times\mathbf{m}_{out}(\mathbf{r})d^3r_i$, where the integral is over the atomic sphere for atom $i$, $\mathbf{B}_{xc,in}(\mathbf{r})$ is the ``input'' exchange-correlation field, which is aligned with the prescribed direction of the magnetization, and $\mathbf{m}_{out}(\mathbf{r})$ the ``output'' magnetization obtained from the NEGF calculation \cite{MacDonald,Freimuth.Bluegel.eaPRB2014,Nikolic,supplement}.
This approach is justified by introducing the constraining fields \cite{Dederichs} stabilizing the instantaneous orientation of magnetization, whereby the internal spin torque is balanced by the torque of the constraining field \cite{supplement}. The spin-density matrix
\begin{equation}
\hat \rho(\mathbf{r}) = -\frac{i}{2\pi} \int\limits_{-\infty}^{\infty} \hat G_<(E,\mathbf{r},\mathbf{r}) dE
\label{dmat}
\end{equation}
is obtained \cite{Faleev} from the Green's function $G_<$ of the Keldysh formalism, given by
\begin{equation}
G_< = i G\left(f_L\Gamma_L+f_R\Gamma_R\right)G^\dagger,
\label{Gless}
\end{equation}
where $G$ and $G^\dagger$ are the retarded and advanced Green's functions, $\Gamma_{L/R}$ is the anti-Hermitian part of the self-energy for lead $L$ (left) or $R$ (right), and $f_{L/R}(E)$ are the occupation functions for the two leads.

The bias $V$ is applied symmetrically, shifting both the potential and the chemical potential of the left (right) lead by $\pm eV/2$. In the steady state of a homogeneous metallic conductor with an applied bias, there is a linear potential drop between the leads, while the density is translationally invariant. Thus, instead of performing a self-consistent calculation for the whole system, we impose a linear potential drop and use equilibrium charge and spin densities for all atoms as inputs in the Kohn-Sham Hamiltonian.

Using the identity $G(\Gamma_L+\Gamma_R)G^\dagger=i(G-G^\dagger)$, the integral in Eq.\ (\ref{dmat}) is formally split in two parts referred to as the Fermi-sea and the Fermi-surface contributions:
\begin{align}
\hat \rho_{sea}(\mathbf{r}) &= \frac{i}{2\pi}\int \bar f(E) (G-G^\dagger) dE\label{sea}\\
\hat \rho_{F}(\mathbf{r}) &= \frac{eV}{4\pi}\int \left(-\frac{\partial \bar f}{\partial E}\right)G(\Gamma_L-\Gamma_R)G^\dagger dE
\label{FS}
\end{align}
where $\bar f$ is the Fermi function with the unperturbed chemical potential, and only the linear term has been kept in (\ref{FS}). This separation is not unique and represents a convenient choice of gauge \cite{gauge}.
In the Fermi-sea contribution (\ref{sea}), the bias enters through the linear potential drop. The Fermi-sea term can contribute to magnetization damping \cite{supplement}.

We consider a Co/Pt bilayer with six monolayers each of Co and Pt. The atoms are placed on the sites of the ideal face-centered cubic (fcc) lattice with the lattice parameter $a=3.75$ \AA, which is approximately half-way between those of fcc Co and Pt. The interface is taken along a (001) plane, and the current direction is [110]. The free surfaces are separated by four monolayers of empty spheres representing vacuum. The length of the active region is 120 monolayers, or 15.9 nm \cite{supplement}.

The thin-film bilayers used for SOT measurements have rather large resistivities in the 20-100 $\mu\Omega\cdot$cm range \cite{Garello.Miron.eaNN2013,Ghosh2017,Safranski.Montoya.eaAe2017}, reflecting a large degree of disorder. The dominant types of defects responsible for the large residual resistivity are not known. As a generic representation, we use the Anderson disorder model, in which a random potential $V_i$ with a uniform distribution in a range $-V_m < V_i < V_m$ is applied on each site $i$, including the empty spheres. In order to gain insight about the mechanisms of SOT and its dependence on the relaxation time $\tau$, we considered four values of $V_m$: 0.77, 1.09, 1.33, and 1.54 eV; the corresponding resistivities range from 23 to 46 $\mu\Omega\cdot$cm \cite{supplement}.

The total torque $\mathbf{T}$ is split into two parts: $\mathbf{T}=\mathbf{T}_e+\mathbf{T}_o$, which are, respectively, even and odd with respect to $\mathbf{m}$. The crystallographic symmetry of the bilayer is $C_{4v}$. We align the $x$ axis with the current direction [110] and $z$ with [001], which is normal to the film plane. Group-theoretical analysis gives the allowed terms in the angular dependence of SOT:
\begin{widetext}
\begin{align}
\mathbf{T}_e&= P(\left\{A\right\},\theta) \;\mathbf{m}\times\left[\left(\mathbf{z}\times\mathbf{E}\right)\times\mathbf{m}\right]+ P(\left\{A^\prime\right\},\theta) \,\left(\mathbf{m}\cdot\mathbf{E}\right)\mathbf{z}\times\mathbf{m}\nonumber\\
&+ P(\left\{A_\alpha\right\},\theta) \, m_z\left(m_x^2-m_y^2\right)\mathbf{m}\times\left(E_x,-E_y,0\right)+ P(\left\{A_\beta\right\},\theta) \, \left[\left(m_x^2-m_y^2\right)(\mathbf{m}\times\mathbf{z})(E_xm_x-E_ym_y)-\left<\dots\right>\right]+\cdots\label{angeven}\\
\mathbf{T}_o&= P(\left\{B\right\},\theta) \;\left(\mathbf{z}\times\mathbf{E}\right)\times\mathbf{m}+ P(\left\{B^\prime\right\},\theta) \;\left(\mathbf{m}\cdot\mathbf{E}\right)\mathbf{m}\times\left(\mathbf{z}\times\mathbf{m}\right)+ P(\left\{B_\alpha\right\},\theta) \,\left(m_x^2-m_y^2\right)\mathbf{m}\times\left(E_y,E_x,0\right)+\cdots
\label{angodd}
\end{align}
\end{widetext}
Here $\left\{X\right\}$ denotes a set of coefficients $X_{2n}$, $n=0,1,\dots$, and $P(\left\{X\right\},\theta)=\sum_{n} X_{2n} P_{2n}(\cos\theta)$ is a linear combination of even Legendre polynomials. The $A$, $A^\prime$, $B$, $B'$ terms are allowed in a system with axial symmetry group $C_{\infty v}$, while the $A_\alpha$, $A_\beta$, $B_\alpha$ terms appear once the symmetry is reduced to $C_{4v}$. $A_0$ and $B_0$ represent the conventional dampinglike and fieldlike SOT terms, respectively.

The brackets $\left<\dots\right>$ in Eq.\ (\ref{angeven}) stand for the average of the preceding term over the axial rotations of the bilayer (which is proportional to a linear combination of $A^\prime$ terms). Such averages already vanish for the $A_\alpha$ and $B_\alpha$ terms. In the axially symmetric polycrystalline sample with (001) texture, the predicted angular dependence is given by the $A$, $A'$, $B$, and $B'$ terms only.

In all calculations we have $\mathbf{E}=E \hat x$, and the torquances are defined as $\boldsymbol{\tau}_e=\mathbf{T}_e/(ME)$, $\boldsymbol{\tau}_o=\mathbf{T}_o/(ME)$, where $M$ is the total magnetization, and have the dimension of a magnetoelectric coefficient $[B/E]=\mathrm{ns/m}=\mathrm{T\cdot nm/V}$.

The contribution of SOT to magnetization damping $\alpha$, which is obtained in ferromagnetic resonance (FMR) linewidth measurements \cite{Safranski.Montoya.eaAe2017}, is $\Delta\alpha=C (E/B)$, where
\begin{align}
 C = \mathbf{m} \cdot \nabla_{\mathbf{m}}\times\left[\mathbf{m}\times\boldsymbol{\tau}(\mathbf{m})\right]
 \label{damping}
\end{align}
is the negative curl of the effective field \cite{supplement}.

The Fermi-sea term is calculated in the middle of the device with a finite bias of order 1 mV applied symmetrically, as required by Eq.\ (\ref{sea})-(\ref{FS}), without disorder. Equilibrium torque from the magnetic anisotropy is removed by subtracting the torque at positive and negative bias. To avoid the formidable task of evaluating the integral in Eq.\ (\ref{sea}), the Fermi-sea term is calculated at a finite temperature, using the integration method of Ref.\ \onlinecite{Ozaki}. The integrand only needs to be calculated at a finite number of points on the imaginary axis, most of which allow a coarse mesh in the reciprocal-space integral. The Fermi-sea term, which is strictly even, is calculated for 61 orientations of the magnetization \cite{points} and then fitted to Eq.\ (\ref{angeven}). We have verified that the Fermi-sea torque depends linearly on the bias voltage, is insensitive to the length of the active region at constant field, and vanishes if the linear potential drop is replaced by two abrupt steps at the edges of the active region.

\begin{figure}[htb]
\includegraphics[width=0.85\columnwidth]{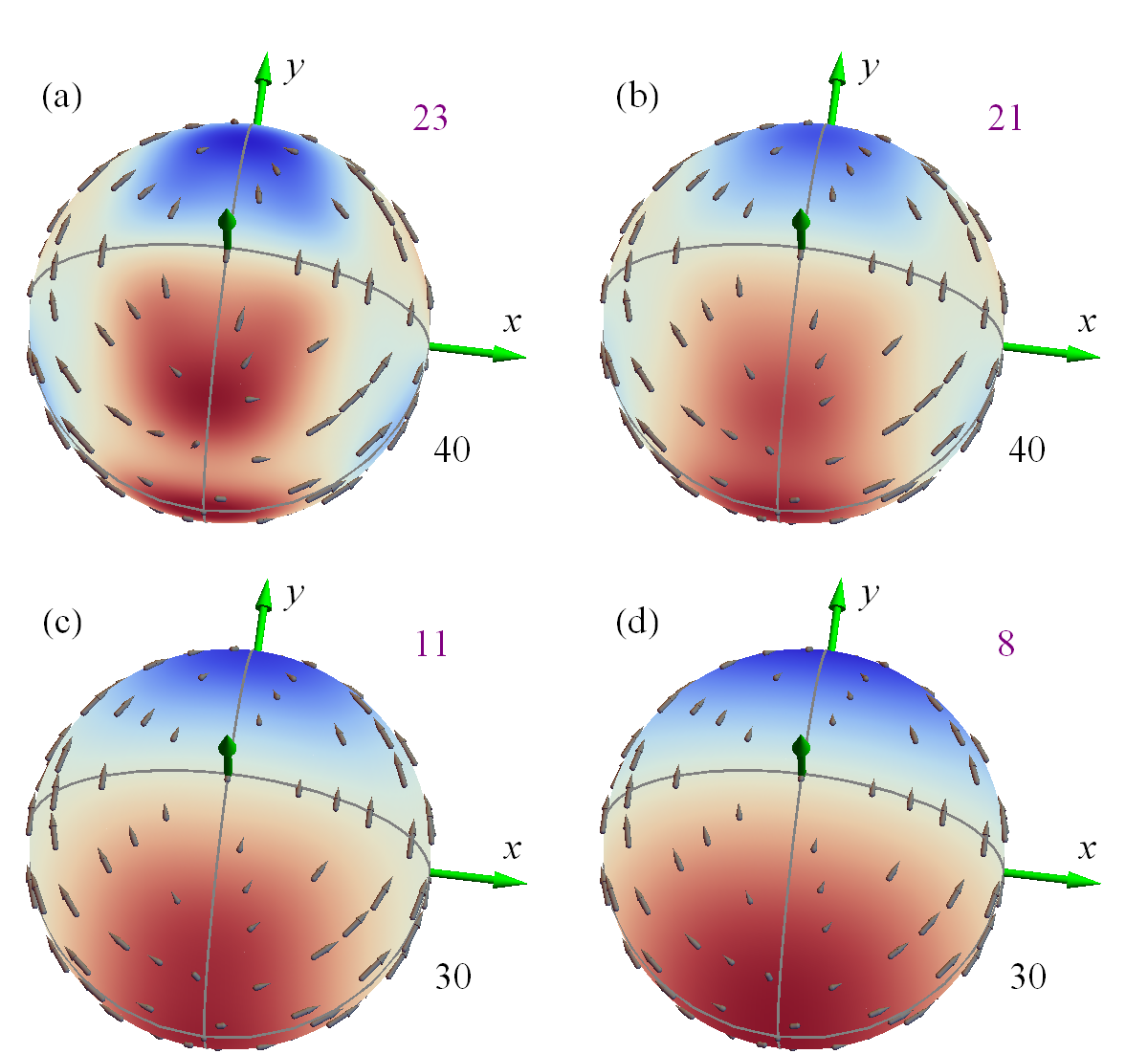}
\caption{Fermi sea contribution to the torquance $\boldsymbol{\tau}_e$ (arrows) at (a) 50 K, (b) 100 K, (c) 200 K, (d) 300 K. The intensity of red (blue) color shows the positive (negative) magnitude of the damping parameter $C$ [Eq.\ (\ref{damping})]. In each panel, the number on bottom right gives the scale of an arrow with a length equal to the sphere radius, and one on top right gives the color map scale (both in ns/m).}
\label{fsea}
\end{figure}

The Fermi-sea torquances obtained for $T=50$, 100, 200, and 300 K are shown in Fig.\ \ref{fsea}. The minimal set of terms giving an acceptable fit at all temperatures includes $A_0$, $A_2$, $A'_0$, and $A_{\beta0}$ (see Table \ref{coefs}); a more accurate multi-parametric fit is used to compute the parameter $C$ shown in Fig.\ \ref{fsea}.
$A'_0$ is the largest term in the minimal fit, and it becomes quite large at low temperatures. $A_2$ and $A_{\beta0}$ are also important at lower $T$, although $A_{\beta0}$ should average out in polycrystalline samples.

\begin{figure}[htb]
\includegraphics[width=0.85\columnwidth]{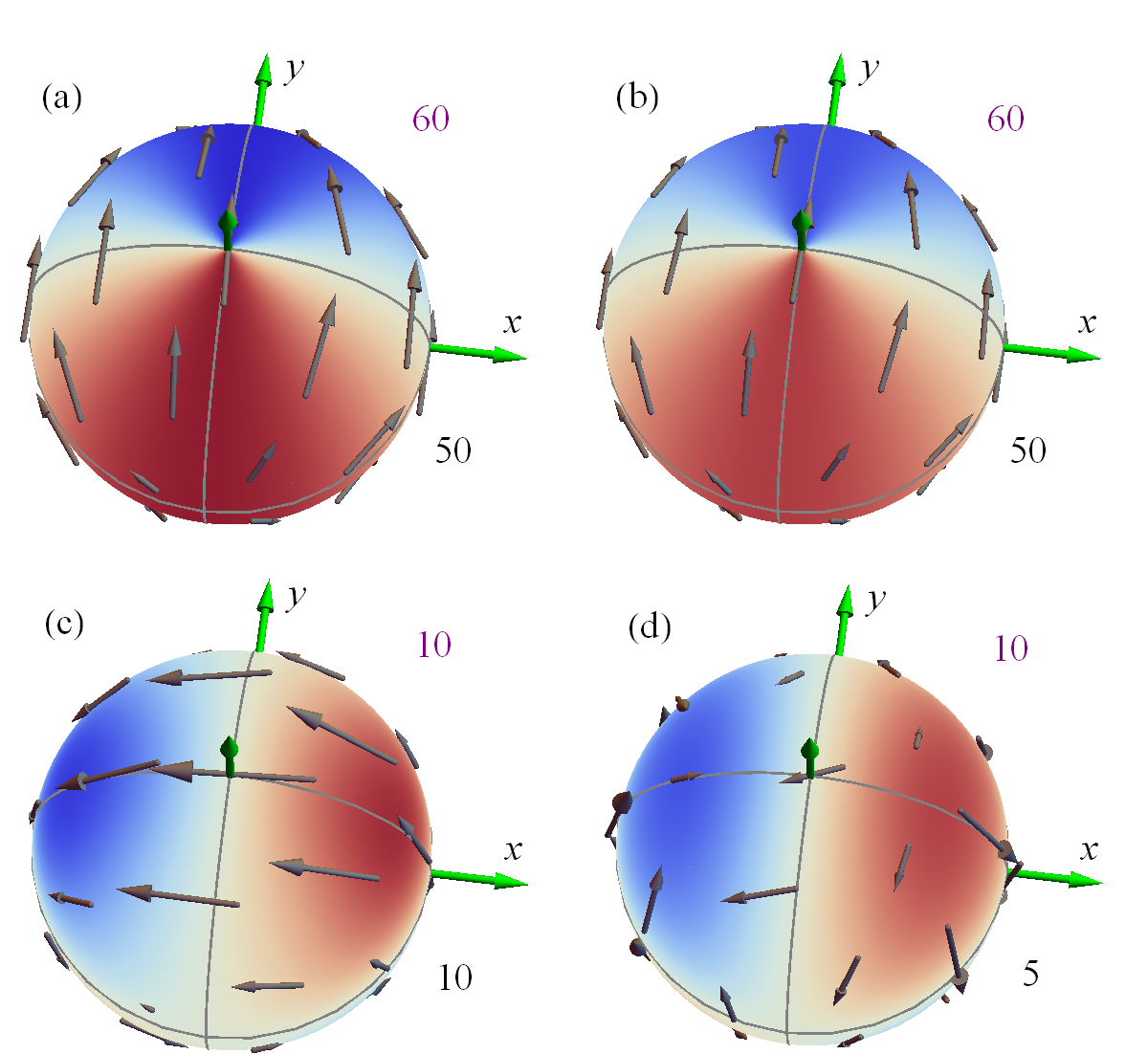}
\caption{Fermi surface contribution to the torquance (arrows): (a) $\boldsymbol{\tau}_e$ at $V_m=0.77$ eV, (b) $\boldsymbol{\tau}_e$ at $V_m=1.54$ eV, (c) $\boldsymbol{\tau}_o$ at $V_m=0.77$ eV, (d) $\boldsymbol{\tau}_o$ at $V_m=1.54$ eV. The scales are indicated as in Fig.\ \ref{fsea}. Supercells with $L_y=2$ were used for disorder averaging.}
\label{fsurf}
\end{figure}

The integrand in Eq.\ (\ref{FS}) for the Fermi-surface term contains a delta-function at zero temperature and needs to be calculated only near the Fermi level $E_F$. The temperature dependence of this term is determined primarily by $\tau$ rather than the temperature in the Fermi distribution function. The Fermi surface contribution to the total torquance, summed up over all sites in the active region, is calculated for 32 orientations of the magnetization, which form 16 antiparallel pairs, and averaged over a sufficient number of disorder configurations \cite{note-averaging}. The symmetric and antisymmetric parts of the torque are then fitted to Eqs.\ (\ref{angeven}) and (\ref{angodd}).
Only $A_0$, $A'_2$, $B_0$, and $B'_0$ coefficients turned out to be sizeable; they are listed in Table \ref{coefs}. With the exception of $A'_0$, all coefficients depend weakly on the transverse supercell size $L_y$, confirming the reliability of disorder averaging. The fitted expressions were used to evaluate the damping parameter $C$, and the results are displayed in Fig.\ \ref{fsurf} for two strengths of disorder, $V_m=0.77$ and 1.54 eV.

The Fermi-surface contribution to the even torquance is dominated by the simple dampinglike term $A_0$. The leading contribution to damping from the even torquance is given by $C=-(2A_0+A'_0)m_y$. Although the Fermi-surface part of $A'_0$ converges slowly with the transverse supercell size $L_y$, it is clear from Table \ref{coefs} that its contribution to $C$ is small compared to $A_0$.

Table \ref{coefs} shows that, as the disorder strength increases from 0.77 to 1.54 eV, the $A_0$ term remains essentially constant, while the resistivity and the resistance of the active region increase by more than a factor of 2 \cite{supplement}. This shows that the dampinglike torquance $A_0$ depends weakly on $\tau$.
The magnitude of $A_0$ is consistent with experimental data \cite{Nguyen} for a Co/Pt bilayer with similar layer thicknesses, as well as with prior calculations using phenomenological broadening \cite{Mahfouzi2018}. These observations suggest that dampinglike SOT in this bilayer is dominated by intrinsic band-structure effects.

\begin{table}[htb]
\caption{Coefficients (ns/m) in the angular expansion of the spin-orbit torquance in the Co/Pt bilayer. $L_y$ is the lateral supercell size in the units of $a/\sqrt{2}$ (only relevant for the Fermi-surface part). $E$ is the energy; $E_\pm=E_F\pm 0.046$ eV.}
\begin{tabular}{|c|c|c|c|c|c|c||c|c|c|c|}
\hline
& \multirow{2}{*}{$E$}& \multirow{2}{*}{$L_y$} & \multicolumn{4}{c||}{Fermi surface, $V_m$ (eV)} & \multicolumn{4}{c|}{Fermi sea, $T$ (K)}\\
\cline{4-11}
& & &                      0.77   &   1.09    &   1.33    &   1.54    & 300 & 200 & 100 & 50 \\
\hline
\multirow{5}{*}{$A_0$}&
$E_F$ &    $1$        &  29.4    &   24.8    &   23.4    &   21.9    &   1.4  &   0.8 & 0.8 & -1.3\\
&$E_F$ &$2$           &  29.9    &   31.3    &   24.4    &   27.7    &      &&&\\
&$E_F$ &$3$           &          &   27.5    &           &           &     &&&\\
&$E_+$ & $2$
        &          &   30.5    &           &           &     &&&\\
&$E_-$ & $2$        &          &   26.8    &           &           &     &&&\\
\hline
\multirow{5}{*}{$A'_0$}&
  $E_F$     & $1$     &  -5.2    &   -3.1    &   -2.6    &   -0.7    &    5.3    &  7.6 & 10.6 & 13.6 \\
&$E_F$ &$2$           &  -3.3    &  -10.7    &   -2.8    &   -7.5    &      & & &\\
&$E_F$ &$3$           &          &   -6.0    &           &           &      & & &\\
&$E_+$ & $2$    &          &   -4.7    &           &           &      & & &\\
&$E_-$ & $2$    &          &   -2.2    &           &           &      & & &\\
\hline
$A_2$ & $E_F$ & $2$    &  -1.3    &   -2.2    &   -0.3    &   -0.9    &   0.6    &  1.4 & 3.2 & 6.3 \\
\hline
$A_{\beta0}$ & $E_F$ & $2$      &  -1.5    &   -0.3    &    0.0      &     0.0       & 0.3    &  1.4 & 5.2 & 7.3 \\
\hline
\multirow{5}{*}{$B_0$}&
$E_F$ &$1$            &  -8.1    &   -8.0    &   -6.3    &   -4.1    & \multicolumn{4}{c|}{}\\
&$E_F$ &$2$           &  -8.8    &   -5.0    &   -3.8    &   -1.7    & \multicolumn{4}{c|}{}\\
&$E_F$ &$3$           &          &   -6.3    &           &           & \multicolumn{4}{c|}{0}\\
&$E_+$ & $2$        &          &   -7.5    &           &       & \multicolumn{4}{c|}{}\\
&$E_-$ & $2$        &          &   -3.2    &           &       & \multicolumn{4}{c|}{}\\
\hline
\multirow{5}{*}{$B'_0$}&
$E_F$ &$1$            &  -6.8    &   -8.2    &   -10.7   &   -9.9    & \multicolumn{4}{c|}{}\\
&$E_F$ &$2$           &  -7.5    &   -7.6    &    -6.8   &   -5.8    & \multicolumn{4}{c|}{}\\
&$E_F$ &$3$           &          &   -8.3    &           &           & \multicolumn{4}{c|}{0}\\
&$E_+$ & $2$        &          &   -9.3    &       &           & \multicolumn{4}{c|}{}\\
&$E_-$ & $2$        &          &   -8.3    &       &           & \multicolumn{4}{c|}{}\\
\hline
\end{tabular}
\label{coefs}
\end{table}

In addition to the simple fieldlike $B_0$ term, the odd torquance contains a sizeable $B'_0$ term of comparable magnitude (see Table \ref{coefs}); other terms are relatively small. This is in contrast to calculations based on phenomenological broadening \cite{Mahfouzi2018}, where no terms beyond $B_0$ were found. The $B_0$ coefficient decreases with increasing disorder strength, as expected for ISGE. However, the relatively large error bar for $B_0$, which is evident from its dependence on $L_y$, does not allow us to predict its temperature dependence at constant current density.

The mechanisms of SOT are closely related to its temperature dependence through their dependence on relaxation time $\tau$. The intrinsic dampinglike SOT is independent of $\tau$ at a fixed electric field, and hence it should be proportional to the resistivity $\rho(T)$ at a constant current density. Although the fieldlike SOT due to interfacial ISGE scales with $\tau$ similar to the conductivity, the interfacial and bulk scattering rates may be different.

There are few experimental measurements of the temperature dependence of SOT, and they are poorly understood. In Ta-based systems the fieldlike SOT was reported to increase quickly with temperature while the resistivity and the dampinglike SOT are nearly constant \cite{Kim2013,Qiu2014}. This behavior is inconsistent with the ISGE mechanism of the fieldlike-SOT. Temperature dependence of the fieldlike SOT is different in as-grown Pt/Co and annealed Pt/CoFeB bilayers \cite{Pai2015}. The unexpected temperature dependence of the fieldlike SOT suggests that processes involving phonons or magnons may play an important role \cite{Manchon.Miron.eaAe2018,Cheng2017}.

The terms $B'_0$ and $B_2$ in the odd torquance contribute to damping as $C=3(B'_0+B_2)m_xm_z$, which is the ``planar Hall-like'' damping observed when $\mathbf{m}$ lies in the $xz$ plane \cite{Safranski.Montoya.eaAe2017}. Table \ref{coefs} shows that the term $B'_0$ is not sensitive to disorder strength, similarly to $A_0$. The $B_2$ term was found to be small in all cases.

The existence of large terms beyond $B_0$ in the odd SOT is consistent with experimental observations \cite{Garello.Miron.eaNN2013,Ghosh2017,Safranski.Montoya.eaAe2017}. However, while we found large $B'_0$ and $B_2\approx0$ in a Co/Pt bilayer, measurements of SOT in AlO$_x$/Co/Pt \cite{Garello.Miron.eaNN2013} and AlO$_x$/Co/Pd \cite{Ghosh2017} suggest an approximate relation $B_2=-\frac23 B'_0$ in these systems \cite{supplement}. The relative magnitude of the damping parameter $C$ measured in the $xy$ (spin-Hall-like SOT) and $xz$ planes (planar Hall-like SOT) agrees with FMR linewidth measurements \cite{Safranski.Montoya.eaAe2017}, but the sign of $B'_0$ is different.
This disagreement may be due the inadequacy of the Anderson disorder model. Indeed, weak dependence of $B'_0$ on disorder strength (see Table \ref{coefs}) and the absence of any terms beyond $B_0$ in calculations based on band broadening \cite{Mahfouzi2018} suggest that these terms arise from vertex corrections, which are sensitive to the type of disorder present in the system.

Table \ref{coefs} also lists the Fermi-surface SOT coefficients calculated at energies $E_\pm=E\pm 0.046$ eV, where $(-\partial\bar f/\partial E)$ is reduced by 50\% compared to its maximal value at 300 K. Weak energy dependence of $A_0$ and $B'_0$, and approximately linear dependence of $B_0$, suggests that these coefficients are not sensitive to the Fermi temperature. The $A'_0$ coefficient remains small.

\begin{figure}[htb]
\includegraphics[width=0.85\columnwidth]{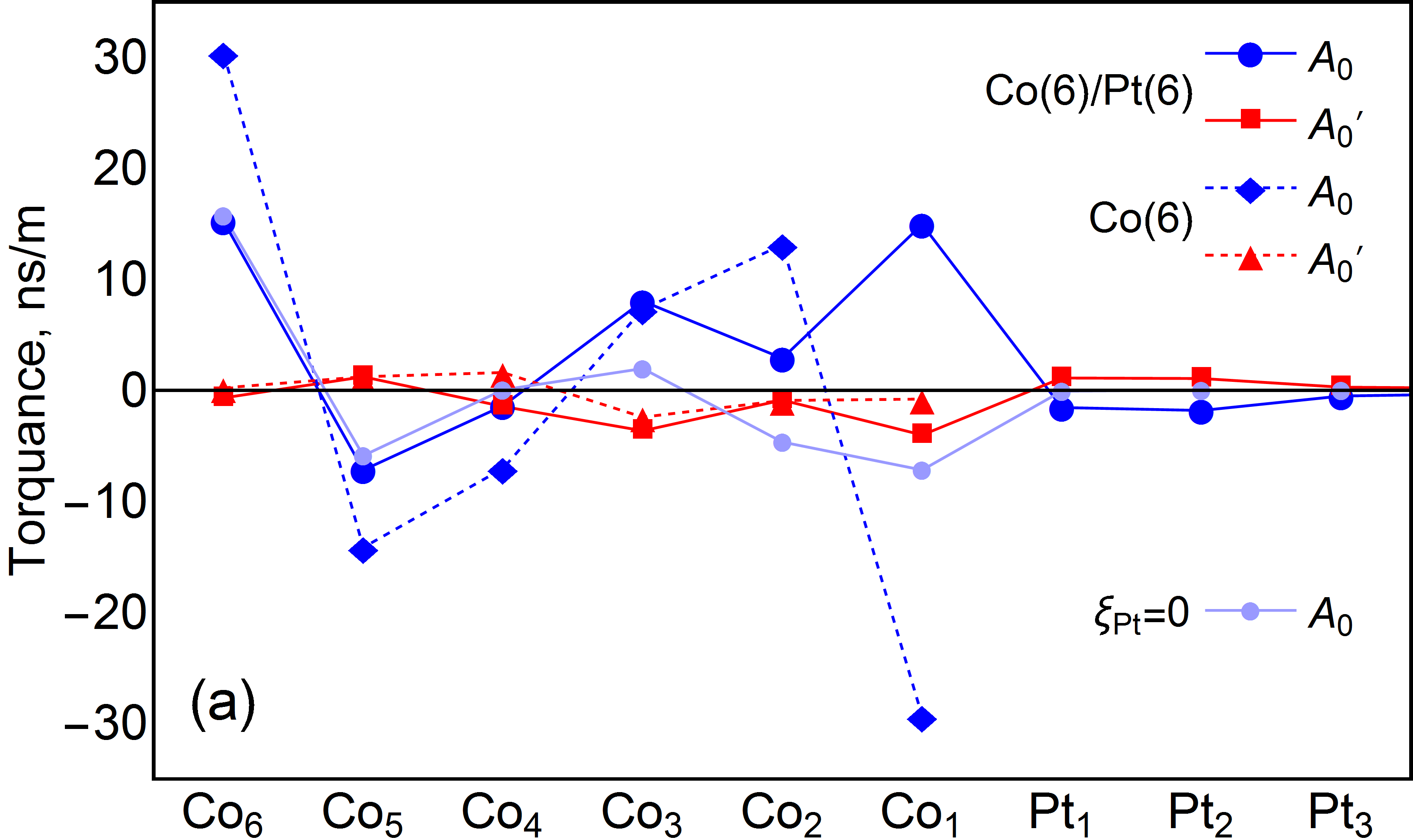}
\vskip1ex
\includegraphics[width=0.85\columnwidth]{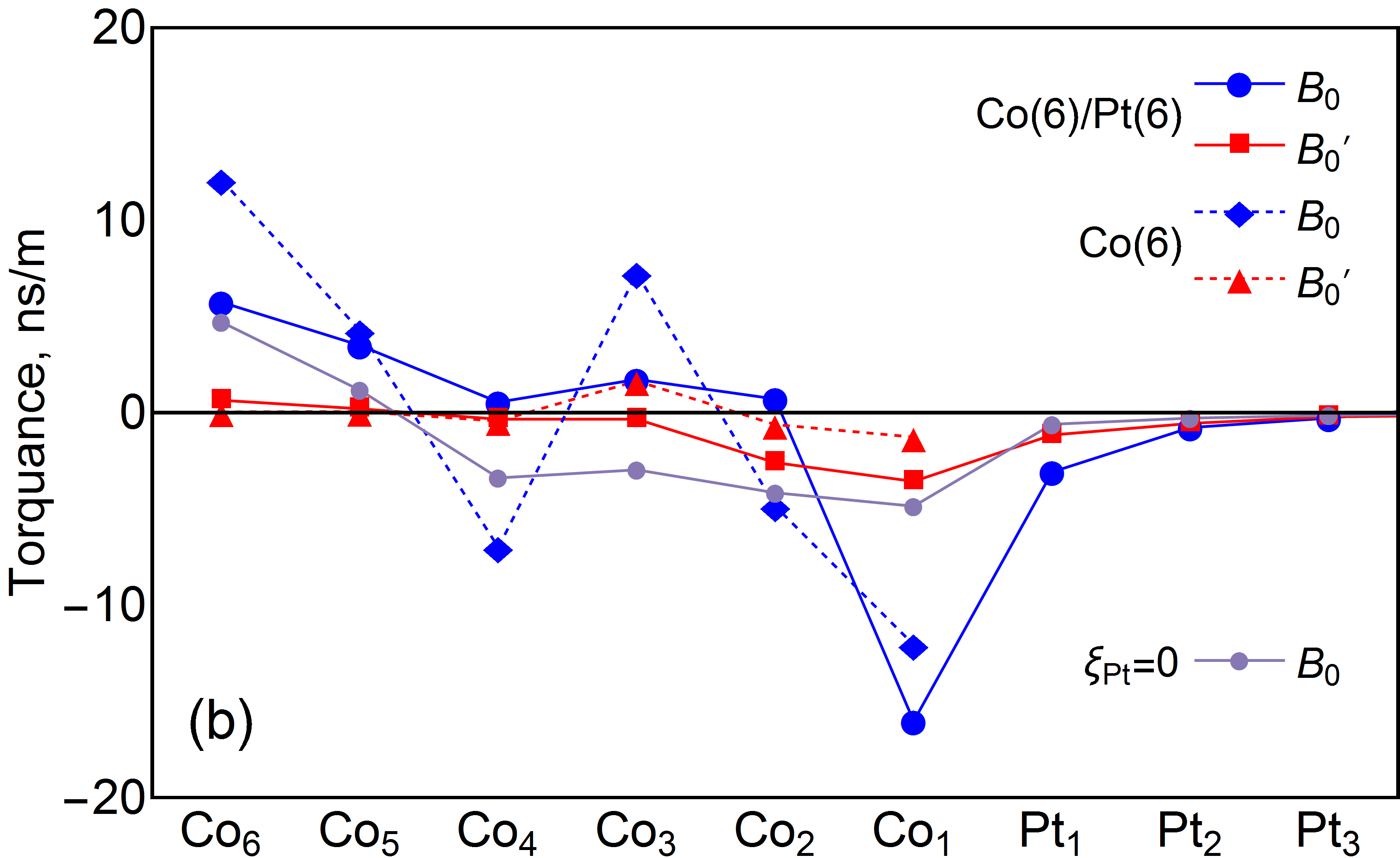}
\caption{Atom-resolved torquances in the Co(6)Pt(6) bilayer (solid lines) and in the free-standing Co(6) film (dashed lines) at $V_m=1.09$ eV, obtained with $L_y=3$. (a) Even terms $A_0$ and $A'_0$, (b) odd terms $B_0$ and $B'_0$. Light-blue curves (labeled $\xi_\mathrm{Pt}=0$): $A_0$ in Co(6)Pt(6) with SOC on Pt atoms set to zero, obtained with $L_y=1$.}
\label{sres}
\end{figure}

For further insight in the origin of SOT, Fig.\ \ref{sres} shows atom-resolved contributions to the $A_0$, $A'_0$, $B_0$, and $B'_0$ terms at $V_m=1.09$ eV. For comparison, these quantities are also shown for the free-standing 6-monolayer Co film with the same lattice parameter, where the total torquance vanishes by symmetry.

The contributions to $A_0$ and $B_0$ are spread throughout the thickness of the film, with the largest contributions coming from the Co atoms at the Co/Pt interface and at the free surface of Co. On the other hand, the $B'_0$ term appears to originate at the Co/Pt interface. It is interesting to observe a considerable contribution to $B_0$ from the Pt atoms near the interface, which carry a magnetic moment of about $0.24\mu_B$ thanks to the magnetic proximity effect \cite{MT}. In fact, SOT on the Pt atoms contributes as much as 40\% of the total magnitude of $B_0$. Surprisingly, the atom-resolved contributions at the surface Co atoms in the free-standing Co film are even larger in magnitude than those at the Co/Pt interface.

Finally, we examine the SOT with the SOC on Pt atoms switched off, using the supercell with $L_y=2$. The $A_0$ term essentially disappears, but, as seen in Fig.\ \ref{sres}(a), atom-resolved contributions remain sizeable, and those near the free Co surface barely change. The $B'_0$ term is strongly suppressed from $-7.6$ to $-1.4$ ns/m, which is comparable to the averaging error. On the other hand, the $B_0$ term increases to $-10.8$ ns/m, with strongly redistributed atom-resolved contributions [Fig.\ \ref{sres}(b)].
These results suggest that, without SOC on Pt, the SOT in our Co/Pt bilayer is nearly non-dissipative, i.e., it does not affect magnetization damping. Current-induced Dzyaloshinskii-Moriya interaction \cite{FreimuthDMI} formally leads to dampinglike atom-resolved torques that add up to zero \cite{supplement}. Thus, strong fieldlike SOT does not require a heavy-metal layer, but understanding the prerequisites for observing dampinglike SOT without heavy metals \cite{Haidar} will require further research.

In conclusion, we have demonstrated the feasibility of calculating the SOT for a Co/Pt bilayer with an explicit model of disorder within the NEGF formalism based on density-functional theory. Terms beyond the usual dampinglike and fieldlike torques were found, including a sizeable planar Hall-like $B'_0$ term [Eq.\ (\ref{angodd})], consistent with FMR measurements \cite{Safranski.Montoya.eaAe2017}. The dissipative part of SOT is almost entirely due to SOC on Pt atoms.

\begin{acknowledgments}
We thank Vladimir Antropov, Gerrit Bauer, Ilya Krivorotov, Farzad Mahfouzi, Branislav Nikoli\'c, Yaroslav Tserkovnyak, and Igor \v Zuti\'c for useful discussions. The work at UNL was supported by the National Science Foundation (NSF) through Grant No.\ DMR-1609776 and the Nebraska Materials Research Science and Engineering Center (MRSEC, Grant No.\ DMR-DMR-1420645). A.K. was supported by the U.S. Department of Energy, Office of Science, Basic Energy Sciences, under Award No. DE-SC0014189. M.v.S. was supported by the EPSRC CCP9 Flagship Project No. EP/M011631/1. Calculations were performed utilizing the Holland Computing Center of the University of Nebraska, which receives support from the Nebraska Research Initiative.
\end{acknowledgments}

\balancecolsandclearpage

\setcounter{figure}{0}
\makeatletter
\renewcommand{\thefigure}{S\arabic{figure}}
\renewcommand{\bibnumfmt}[1]{[S#1]}
\renewcommand{\citenumfont}[1]{S#1}

\section*{Supplemental Material}

\subsection{Calculation of spin torques in density functional theory}

In first-principles calculations based on density functional theory (DFT), the non-equilibrium spin torque on atom $i$ is usually calculated \cite{SMacDonald,SFreimuth,SNikolic} as
\begin{equation}
\mathbf{T}_i=\int\mathbf{B}_{xc,in}(\mathbf{r})\times \mathbf{m}_{out}(\mathbf{r})\,d^3r_i,
\label{inout}
\end{equation}
here $\mathbf{B}_{xc,in}(\mathbf{r})$ is the exchange-correlation field used to construct the Kohn-Sham potential, and $\mathbf{m}_{out}(\mathbf{r})$ the magnetization density obtained as the \emph{output} from this potential in the presence of an applied bias.
A similar expression was derived in multiple scattering formalism \cite{STurzh,SAntropovMRS} to represent the variation of the total energy with respect to the rotation of a magnetic moment.
According to these formulas, the spin torque appears due to the misalignment between the orientation of the exchange-correlation field $\mathbf{B}_{xc}$, which represents the assumed ``input'' direction of the magnetic moment, and the calculated ``output'' spin moments. Because a magnetic configuration with such misalignment is not self-consistent, rigorous consideration in the static case requires the introduction of auxiliary external fields within the constrained DFT \cite{SDederichs}. Mapping of the constrained-DFT total energy surface to an effective spin Hamiltonian leads to adiabatic spin dynamics equations \cite{SAntropov,SHalilov} with the corresponding effective fields. However, dampinglike torque in the non-equilibrium case can not be described through such mapping. Here we show that similar considerations using constraining fields can be used to justify Eq.\ (\ref{inout}) in this case.

The appropriate framework in the dynamical case is given by time-dependent DFT (TD-DFT) \cite{SRunge}. The evolution of the spin density matrix is given by its commutator with the Kohn-Sham Hamiltonian, but Eq.\ (\ref{inout}) does not immediately follow. Indeed, TD-DFT contains no reference to the ``output'' spin density; the spin density operator commutes neither with the kinetic energy nor with the spin-orbit coupling operator; and, in the adiabatic local density approximation (ALDA), the exchange-correlation field is everywhere collinear with the spin density and, therefore, exerts no torque on it \cite{SCapelle}.

The TD-DFT equation of motion for a local spin moment on atom $i$ is \cite{SAntropov,SCapelle}
\begin{align}\label{eom}
  \dot{\mathbf{s}}_i &= -\int\langle\hat{J}_s\rangle d\mathbf{S}_i + i\left<\left[\hat{\mathbf{s}}_i,\hat V_\mathrm{SO}\right]\right> \nonumber\\
  &+ \int\mathbf{m}(\mathbf{r})\times\left[\mathbf{B}_{xc}(\mathbf{r})+\mathbf{B}_{ext}(\mathbf{r})\right]d^3r_i,
\end{align}
where $\hat{J}_s=(-i/4)\hat{\boldsymbol{\sigma}}\otimes\bigl(\vec\nabla-\cev\nabla\bigr)$ is the spin current operator, the integrals are over the Wigner-Seitz cell for site $i$ (or over the atomic sphere in the LMTO method), $\hat V_\mathrm{SO}$ is the spin-orbit coupling operator, $\mathbf{m}(\mathbf{r})=\frac{\gamma}{2}\phi^*_\mathrm{KS}\hat{\boldsymbol{\sigma}}\phi_\mathrm{KS}$ the magnetization density corresponding to the Kohn-Sham wave function $\phi_\mathrm{KS}$, $\mathbf{B}_{xc}(\mathbf{r})$ the exchange-correlation field, and $\mathbf{B}_{ext}(\mathbf{r})$ the external field. All expectation values are taken over $\phi_\mathrm{KS}$.
Because we work within ALDA, the vectors $\mathbf{B}_{xc}(\mathbf{r})$ and $\mathbf{m}(\mathbf{r})$ are everywhere collinear, and the $\mathbf{m}(\mathbf{r})\times\mathbf{B}_{xc}(\mathbf{r})$ term in Eq.\ (\ref{eom}) vanishes identically.

The first term in the right-hand side of (\ref{eom2}) can have a longitudinal part which we ignore, assuming that spin dynamics is sufficiently slow so that the magnitudes of $\mathbf{s}_i$ adiabatically follow their directions. Designating the first two terms in Eq.\ (\ref{eom}) as the internal torque $\mathbf{T}_{i,int}$, and the torque of the external field as $\mathbf{T}_{i,ext}$, we have
\begin{equation}\label{eom2}
  \dot{\mathbf{s}}_i = \mathbf{T}_{i,int} + \mathbf{T}_{i,ext}.
\end{equation}
Our goal is to find $\mathbf{T}_{i,int}$, generally in the presence of a non-equilibrium spin density induced by the electric field. Note that $\mathbf{T}_{i,int}$ is given by the sum of the influx of the transversely polarized spin current and the on-site spin-orbit torque. In particular, in a noncollinear magnetic state the spin-current term includes the torque produced by the conventional exchange interaction.

Consider some, generally noncollinear, spin configuration specified by a set of ``input'' spin orientations $\hat s_{i,in}$. In the DFT calculation, these orientations are generally not self-consistent: the ``output'' spin orientations $\hat s_{i,out}$ from the solution of the Kohn-Sham equations are not parallel to $\hat s_{i,in}$ (which prescribe the directions of the exchange-correlation fields). This misalignment is associated with the emergence of spin torques $\mathbf{T}_{i,int}$ in TD-DFT. If the local moment on a given site is sufficiently large, the misalignment tends to be small.

Instead of calculating the torques $\mathbf{T}_{i,int}$ explicitly from Eq.\ (\ref{eom}), we observe that the given instantaneous spin orientations $\hat s_i$ can be stabilized by transverse constraining fields $\mathbf{B}_{ci}$ chosen so that $\hat s_{i,out}=\hat s_{i,in}$. With these constraining fields the spin configuration is self-consistent, and the total torques vanish.

As is common in constrained DFT calculations \cite{SStocks}, let us choose the constraining fields with the same radial dependence as the exchange-correlation field: $\mathbf{B}_{ci}(\mathbf{r})=B_{i,xc}(r)\mathbf{t}_i$, where $\mathbf{t}_i \cdot \hat{s}_i=0$. The addition of such constraining fields is equivalent to a rigid rotation of the ``input'' exchange-correlation fields by small angles. For atoms with large spin moments, it is a good approximation to assume that the ``output'' spin densities also rotate rigidly by the same angles. Because the constrained state is stationary, it is then easy to see that
\begin{align}
\mathbf{T}_{i,int} = - \mathbf{T}_{i,ext} &= - \int \mathbf{m}(\mathbf{r})\times \mathbf{B}_{ci}(\mathbf{r})d^3r_i\nonumber\\
&= \int \mathbf{B}_{xc,in}(\mathbf{r})\times \mathbf{m}_{out}(\mathbf{r})d^3r_i,
\label{torque}
\end{align}
and we return to Eq.\ (\ref{inout}).

\subsection{Fermi sea contribution to the dampinglike torque}
As it was shown in the main text, an explicit calculation for a metallic bilayer yields a finite Fermi-sea contribution in the middle of the active region, which contributes to magnetization damping. At first sight, this result seems to disagree with the conclusion of Ref.\ \cite{SMahfouzi16} that the Fermi-sea contribution to the dampinglike torque vanishes. This apparent contradiction is resolved by recognizing that the result of Ref.\ \cite{SMahfouzi16} applies only to the \emph{total} torque acting on a finite magnetic nanostructure, while the \emph{local} contribution to the dampinglike torque can be finite. In the NEGF calculation for a long metallic system with a potential gradient, both Fermi-sea and Fermi-surface contributions to the dampinglike torque are finite. However, only the total torque is extensive and uniform. Taken in isolation, the Fermi-sea and Fermi-surface contributions are, therefore, not physically meaningful.

The Fermi sea contribution is calculated for a system with a linear potential drop but assuming a constant electrochemical potential. The torque at site $i$ can be found from the retarded Green's function $\hat G^R$:
\begin{equation}
\mathbf{T}_i = -\frac{1}{2\pi}\mathrm{Im}\mathrm{Tr}\int B_{xc}(r)\mathbf{m}\times\hat{\pmb{\sigma}} \hat G^R(\mathbf{r},\mathbf{r}) dE d^3r_i,
\label{Ti}
\end{equation}
where $\mathbf{m}$ is a unit vector in the direction of the uniform magnetization, the integral is taken over the atomic sphere (or Wigner-Seitz cell) of site $i$, the trace is over the spin indices, and $B_{xc}$, which is spherically symmetric in the atomic sphere approximation, is in energy units. The Green's function is a function of energy $E$ and of $\mathbf{m}$. We also define the effective field $\mathbf{B}_i$ (in energy units):
\begin{equation}
\mathbf{B}_i = \mathbf{T}_i\times \mathbf{m}.
\label{Bi}
\end{equation}

The contribution of the spin torque to magnetization damping, which can be found from the Zeeman energy loss for an infinitesimal magnetization precession loop, is proportional to the negative curl of the effective field:
\begin{equation}
\Delta\alpha = -\frac{1}{E_Z}\left[\mathbf{m}\times\nabla_\mathbf{m}\right]\cdot\sum_i\mathbf{B}_i,
\label{curl}
\end{equation}
where $E_Z=MB$ is the total Zeeman energy of the sample in the external field that drives the precession. Eq.\ (\ref{curl}) is general and not limited to the Fermi-sea part.

Let us denote $c_i=-\left[\mathbf{m}\times\nabla_\mathbf{m}\right]\cdot \mathbf{B}_i$.
Using the vector identity $\nabla\times u\mathbf{a}=u\nabla\times\mathbf{a}-\mathbf{a}\times\nabla u$, we find
\begin{equation}
c_i = \frac{1}{2\pi}\mathrm{Im}\,\mathrm{Tr}\int B_{xc}(r)(\hat{\pmb{\sigma}}\times\mathbf{m})\cdot \nabla_\mathbf{m}\hat G^R(\mathbf{r},\mathbf{r}) dE d^3r_i.
\label{Ci}
\end{equation}

Because a small rotation of $\mathbf{m}$ perturbs the Hamiltonian as $\delta H=-\frac12B_{xc}(r)\hat{\boldsymbol{\sigma}}\delta\mathbf{m}$, we have:
\begin{equation}
\nabla_\mathbf{m}\hat G^R(\mathbf{r},\mathbf{r})=-\frac12 \int \hat G^R(\mathbf{r},\mathbf{r}')B_{xc}(\mathbf{r}')\hat{\pmb{\sigma}}\hat G^R(\mathbf{r}',\mathbf{r})d^3r'.
\end{equation}
Inserting this into (\ref{Ci}), we obtain the well-known expression for the (anisotropic) exchange interaction with vertices $(\mathbf{m}\times\hat{\pmb{\sigma}})$ and $\hat{\pmb{\sigma}}$. In order to sort out the Cartesian indices, let us align the spin quantization axis $\hat z$ parallel to $\mathbf{m}$. Then we obtain
\begin{equation}
c_i=-\sum_j (J^{xy}_{ij}-J^{yx}_{ij}),
\label{Ci2}
\end{equation}
where the exchange parameters
\begin{align}
J^{\alpha\beta}_{ij}=-&\frac{1}{2\pi}\mathrm{Im}\;\mathrm{Tr}\int dE d^3r_i d^3r_i\nonumber\\
&\hat\sigma_\alpha B_{xc}(\mathbf{r}_i)\hat G^R(\mathbf{r}_i,\mathbf{r}_j)\hat\sigma_\beta B_{xc}(\mathbf{r}_j)\hat G^R(\mathbf{r}_j,\mathbf{r}_i)
\end{align}

The antisymmetric part of the exchange interaction in (\ref{Ci2}) is the $z$ component of the Dzyaloshinskii-Moriya (DM) vector $\mathbf{D}_{ij}$. Given that the $z$ axis was chosen along $\mathbf{m}$, we can write, in a frame-independent form:
\begin{equation}
c_i = -\mathbf{m}\sum_{j}\mathbf{D}_{ij} \equiv \mathbf{m}\mathbf{D}_{0i}.
\label{Ci3}
\end{equation}
Because $\Delta\alpha$ in (\ref{curl}) is proportional to $\sum_i c_i$, and, by definition, $\mathbf{D}_{ij}=-\mathbf{D}_{ji}$, the total Fermi-sea contribution to damping vanishes in any finite system, in agreement with the conclusion of Ref.\ \onlinecite{SMahfouzi16}.
This result also holds if the system is periodic in all dimensions in which it is infinite, such as a bilayer with an arbitrary periodic potential profile.

On the other hand, the \emph{local} quantities $\mathbf{D}_{0i}$ do not vanish. Due to the presence of a potential gradient, the Fermi-sea calculation with a constant electrochemical potential refers to an \emph{inhomogeneous} system. The sum over the sites $i$ filling the cross-section of the bilayer should give an effective axial DM vector compatible with the symmetry of the system. In the simplest case of the $C_{\infty v}$ symmetry, the term $\mathbf{D}\propto\mathbf{z}\times\mathbf{E}$ is allowed, which, according to (\ref{Ci3}), gives a contribution to damping proportional to $m_y$. This corresponds to the dampinglike torque $\propto \mathbf{m}\times(\mathbf{y}\times\mathbf{m})$. Terms with more complicated angular dependence in Eq.\ (5) of the main text are also allowed.

In an actual device, a finite magnetic bilayer is attached to non-magnetic leads. The Fermi-sea contribution to the total torque is strictly non-dissipative: the edges exactly cancel the bulk contribution. The Fermi-sea contribution from the edges scales with the length of the sample, because the difference between the DM vectors at the two edges scales with this length. On the other hand, the total physical torque at the edges must be finite. Therefore, the divergent Fermi-sea contribution from the edges must be cancelled by the similarly divergent Fermi-surface contribution from those edges. In addition, neither the Fermi-sea nor the Fermi-surface contributions are uniform. Thus, only the sum of the Fermi-sea and Fermi-surface contributions has a clear physical meaning.

If the metallic bilayer is sufficiently long, the total torque is extensive: it scales with the length, while the edge effects are finite. Therefore, we are justified in computing the torque per unit length in the middle of a bilayer attached to infinite magnetic leads. In this calculation, both Fermi-surface and Fermi-sea contributions to the dampinglike torque are finite.

Let us, finally, consider the bulk of an insulator, far from the edges, in a homogeneous electric field. At $T=0$ the placement of the chemical potential inside the band gap has no effect on the charge and spin densities. Therefore, the DM vectors $\mathbf{D}_{ij}$ are translationally invariant, and the mesoscopic average $\langle\mathbf{D}_{0i}\rangle$ vanishes. Therefore, there can be no dampinglike torque in the bulk of an insulator in a homogeneous electric field.

\subsection{Angular dependence of the effective fields}

The even Fermi-surface torque is dominated by the simple antidamping term, which does not require any additional discussion. Here we make a connection with the experimental measurements of the angular dependence of the odd spin-orbit torque \cite{SGarello.Miron.eaNN2013}. In our calculations for the Co/Pt bilayer, this torque is dominated by the terms $B_0$ and $B'_0$ in Eq.\ (6) of the main text:
\begin{equation}
\mathbf{T}_o= B_0\left(\mathbf{z}\times\mathbf{E}\right)\times\mathbf{m}+ B'_0 \left(\mathbf{m}\cdot\mathbf{E}\right)\mathbf{m}\times\left(\mathbf{z}\times\mathbf{m}\right)
\label{To}
\end{equation}
The effective field $\mathbf{B}_e=\mathbf{T}_o\times\mathbf{m}$ corresponding to this torque is referred to in Ref.\ \onlinecite{SGarello.Miron.eaNN2013} as $\mathbf{B}^\perp$ and in Ref.\ \onlinecite{SGhosh2017} as $\mathbf{B}^\mathrm{FL}$ \cite{NoteFL}. Retaining for the moment the term $B_2$ in the torque, we obtain:
\begin{align}
\mathbf{B}_e=-B_e^\theta\cos\theta\sin\phi\mathbf{e}_\theta-B_e^\phi\cos\phi\mathbf{e}_\phi
\label{Be}
\end{align}
where
\begin{align}
B_e^\theta&=B_0+B_2 P_2(\cos\theta)\nonumber\\
B_e^\phi&=B_0-\frac23 B'_0+\left(B_2+\frac23B'_0\right)P_2(\cos\theta).
\label{Beff}
\end{align}
Eq.\ (\ref{Be}) is equivalent to Eqs.\ (4) of Ref.\ \onlinecite{SGhosh2017} where these coefficients are denoted $B_\theta^\mathrm{FL}$ and $B_\phi^\mathrm{FL}$ \cite{NoteFL}.

It was reported in Refs.\ \onlinecite{SGarello.Miron.eaNN2013,SGhosh2017} for Al$_x$O/Co/Pt and Al$_x$O/Co/Pd systems that $B_e^\phi$ in these systems depends weakly on $\theta$. This implies, in our notation, an accidental relation $B_2=-\frac23 B'_0$ that is not required by any symmetry. As seen from Table I in the main text, our calculations predict that $B_2$ is small while $B'_0$ is large, which implies that $B_e^\phi$ depends strongly on $\theta$ while $B_e^\theta$ does not. However, as mentioned in the main text, the $B'_0$ term may be sensitive to the type of disorder present in the system, in which case the Anderson model may not be an adequate representation of the devices studied in Refs.\ \onlinecite{SGarello.Miron.eaNN2013,SGhosh2017}.

The $B'_0$ and $B_2$ terms contribute in exactly the same way to the angular dependence of damping and, therefore, can not be distinguished in the FMR linewidth measurements.

\subsection{Resistivity of the C\lowercase{o}/P\lowercase{t} bilayers}

\begin{figure}[ht]
\includegraphics[width=0.9\columnwidth]{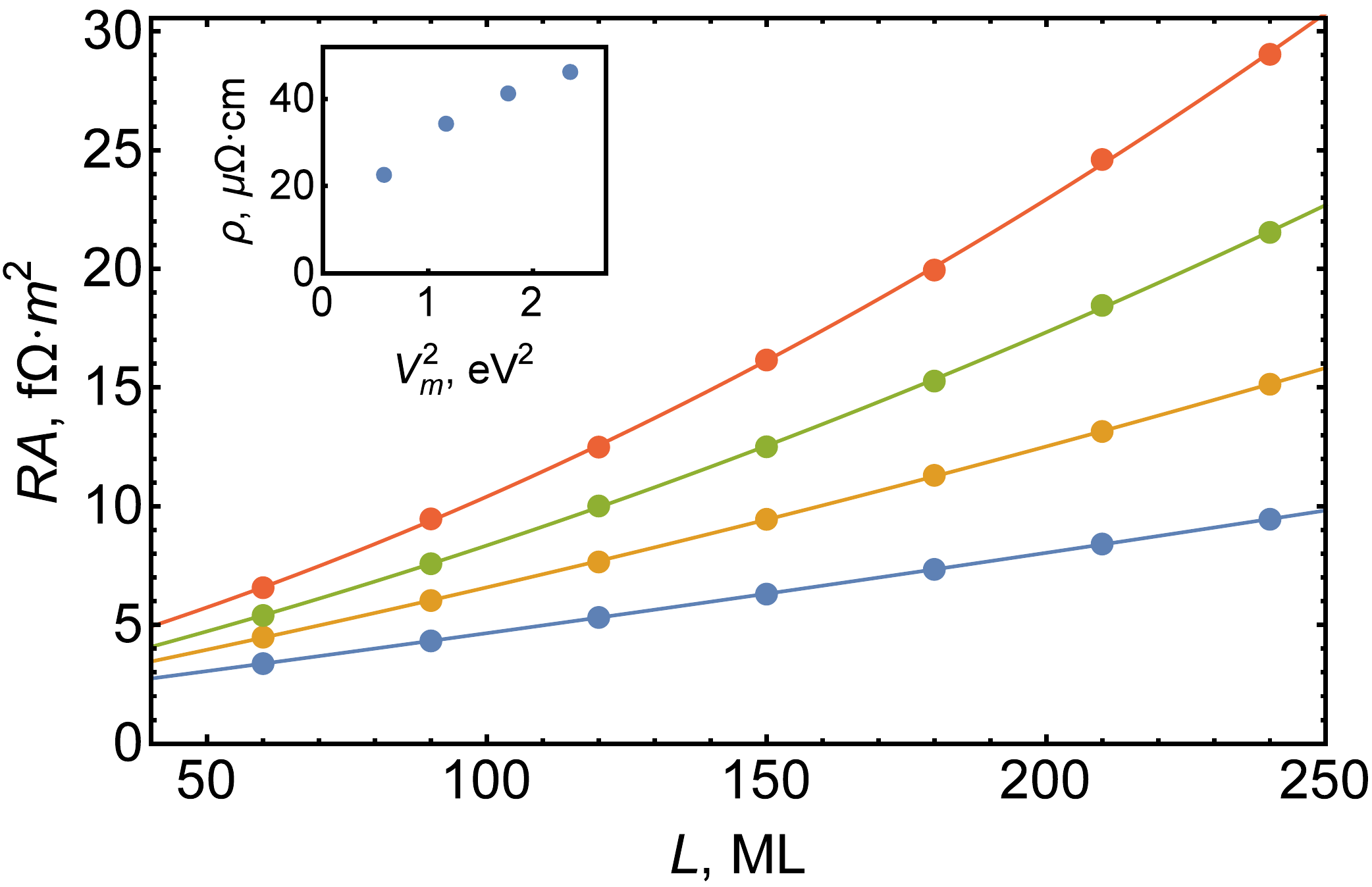}
\caption{Resistance-area product $RA$ as a function of the length of the active region. The four data sets correspond to disorder strengths $V_m=0.77$, 1.09, 1.33, and 1.54 eV. Each curve is a quadratic fit to the corresponding data set. Inset: Resistivity $\rho$ as a function of $V_m^2$, obtained as a linear term from the quadratic fit.}
\label{resist}
\end{figure}

Figure \ref{resist} shows the calculated resistance-area product, as a function of length, for four values of $V_m$. Each data point was obtained by averaging the conductance, calculated with spin-orbit coupling included and the magnetization oriented perpendicular to the plane of the bilayer, over 50 disorder configurations, without enlarging the unit vector in the $y$ direction. The data fit very well to a quadratic function, with positive deviations from linearity arising due to Anderson localization in quasi-one-dimensional supercells. The effective resistivity, estimated from the linear term in the fit, varies between 23 and 46 $\mu\Omega\cdot$cm.

\begin{thebibliography}{99}
\bibitem{note-averaging} Because the terms in Eq.\ (\ref{FS}) with $\Gamma_L$ and $\Gamma_R$ give identical contributions to the odd torquance, it is advantageous to average these terms independently over different disorder configurations.
\end{thebibliography}
\begin{thebibliography}{9}
\bibitem{NoteFL} We avoid referring to the odd torque as ``field-like'' because $B'_0$ and higher-order terms contribute to magnetization damping.
\end{thebibliography}
\end{document}